\def\<{\langle}
\def\>{\rangle}

\documentclass{Rinton-P9x6}
\begin{document}
\title{Quantum Metrology: Detection of weak forces using Schr\"odinger Cat resources}

\author{K.~ Nemoto$^\dag$, W. J. Munro$^\ddag$, G.~J.~Milburn$^*$
and S. L. Braunstein$^\dag$}

\address{$^\dag$ School of Informatics, Dean Street, University of Wales,
Bangor LL57 1UT, UK\\
$^\ddag$ Hewlett-Packard Laboratories, Filton Road, Stoke
Gifford, Bristol BS34 8QZ, UK\\
$^*$ Centre for Quantum Computer Technology, University of Queensland,
Brisbane, QLD, Australia  }


\maketitle

\abstracts{We investigate the utility of  non classical states of
simple harmonic oscillators (a superposition of coherent states) 
for sensitive force detection.  We find that like squeezed states
a superposition of coherent states allows the detection of displacement measurements at
the Heisenberg limit. Entangling many superpositions of coherent states offers a significant
advantage over a single mode superposition states with the same mean photon number.}

\section{Introduction}
Non classical states of light have received considerable attention in the field
of quantum and atom optics. Many non-classical states of light  have been
experimentally produced and characterised. These
states include photon number states, squeezed states and certain entangled
states. There are a number of suggested, and actual, applications of these states
in quantum information processing including; quantum computation, communication and 
cryptography to name but a few. They have also been proposed for high 
precision measurements which include improving the sensitivity of Ramsey fringe
interferometry\cite{Huelga} and the detection of weak tidal forces due 
to gravitational radiation. In this article we consider how non classical states
of simple harmonic oscillators may be used to improve the detection
sensitivity of weak classical forces.

Let us begin by establishing the classical limit for force detection. When 
a classical force, $F(t)$, acts  for a fixed
time on a simple harmonic oscillator, it displaces
the complex amplitude of the oscillator in phase
space with the amplitude and phase of the
displacement determined by the time dependence of the force\cite{Braginsky92}.
The action of the force in the interaction picture is simply 
represented by the unitary displacement operator
\begin{equation}
D(\alpha)=\exp(\alpha a^\dagger -\alpha^* a) \nonumber
\end{equation}
where $a,a^\dagger$ are the annihilation and
creation operators for the single mode
of the electromagnetic field satisfying $[a,a^\dagger]=1$,
and $\alpha$ is a complex amplitude which
determines the average field amplitude, $\langle a\rangle = \alpha$. For
simplicity we will assume that the force displaces the
oscillator in a phase space direction that is orthogonal to the
coherent amplitude of the initial state. To detect the force we would need to measure 
the signal of the quadrature $\hat{Y}=-i(a-a^\dagger)$. If the
oscillator begins in a coherent state $|\alpha_0\rangle$,
($\alpha_0$ is real) the displacement
$D(i\epsilon)$ causes the coherent state to evolve to
$e^{i \epsilon \alpha_0}|\alpha_0+i\epsilon\rangle$. The signal to noise ratio 
is then $SNR=S/\sqrt{V}=2\epsilon$ which must be greater than unity 
to be resolved. This establishes the standard quantum limit (SQL)\cite{WallsMil94} of 
$\epsilon_{SQL}\geq 1/2$.

It is well known\cite{Caves} that this limit may be overcome if the 
oscillator is first prepared in a squeezed state 
(a uniquely quantum mechanical state) for which the uncertainty in the 
momentum quadrature is reduced below the coherent state level. For a momentum squeezed state  
(with mean photon number $n_{tot}$) it is straightforward to show the minimum detectable displacement is
$\epsilon_{min}\geq 1/\sqrt{4 n_{tot}}$. This scales as the inverse square root of the mean photon number 
and means the only way to improve the sensitivity is to increase $n_{tot}$. Can we do better using 
different non-classical resources.

\section{Weak force detection utilising Schr\"odinger cat states.}

\begin{figure}[!bht]
\center{ \epsfxsize=21pc \epsfbox{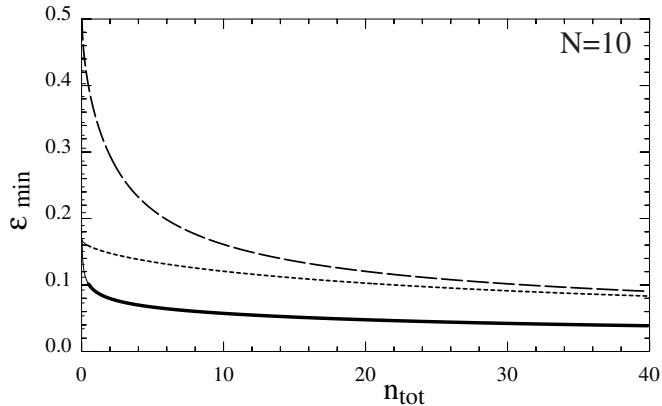}
\caption{Plot of the minimum detectable displacement $e_{min}$ versus $n_{tot}$ for $N=10$. The solid curve 
is for the entangled cat state (\ref{nmodecat}) while $N$ separable copies of a single mode cat state 
$|\alpha\rangle+|-\alpha\rangle$ are given by the dotted line. The final curve (the dashed line) represents 
a single mode cat state with mean photon number $n_{tot}$.}}
\label{fig1}
\end{figure}

We now consider another class of non classical states, based on 
a coherent superposition of coherent states, which can be entangled over $N$ modes.
The $N$ mode entangled cat state has the form
\begin{equation}\label{nmodecat}
|\psi\rangle=\frac{1}{\sqrt{2}}(|\alpha,\alpha,\ldots, \alpha\rangle+
|-\alpha,-\alpha,\ldots,-\alpha \rangle)
\end{equation}
where $n_{tot}$ is the total photon number of the entire state.
Parkins and Larsabal\cite{Parkins2000} recently suggested how this highly
entangled state might be formed in the context of cavity QED and quantised
motion of a trapped atom or ion. The weak force 
$D(i \epsilon)$ (for $\epsilon \ll 1$) acts simultaneously on all modes of 
this $N$ party entangled state. The resulting state is
\begin{eqnarray}\label{fulldisplaced}
|\psi (\theta)\rangle&=&\frac{e^{i N \theta}}{\sqrt{2}} |\alpha+i \epsilon,\ldots,\alpha+i \epsilon\rangle  + 
\frac{e^{-i N \theta}}{\sqrt{2}}|-\alpha+i \epsilon,\ldots,-\alpha+i \epsilon \rangle \nonumber
\end{eqnarray}
where $\theta= \epsilon \alpha$. The theory of optimal parameter estimation indicates that the
limit on the precision with which the displacement parameter  be estimated is bounded by
\begin{eqnarray}
\epsilon_{min}&=&\frac{1}{\sqrt{N \left[1+4 N \alpha^2/(1+e^{-2 N \alpha^2})\right]}}
\;\;{\rm where}\;\;
n_{tot}=N\alpha^2 \frac{\left[1-e^{-2 N \alpha^2}\right]}{\left[1+e^{-2 N \alpha^2}\right]}
\label{ent_emin}
\end{eqnarray}
and realises the Heisenberg Limit. For $n_{tot}\gg 1$ Eqn (\ref{ent_emin}) simplifies to
\begin{eqnarray}
\epsilon_{min}&=&\frac{1}{\sqrt{N \left[1+4 n_{tot}\right]}}\sim \frac{1}{\sqrt{4 N n_{tot}}} \nonumber
\end{eqnarray}
Let us now compare this to the case where one has $N$ copies of the single mode cat state
$|\alpha\rangle+|-\alpha\rangle$. In this case (for $\alpha^2 \gg 1$) the minimum detectable displacement is
$\epsilon_{min}=1/{\sqrt{\left[N+4 n_{tot}\right]}}$ while a single mode cat state 
with mean photon number $n_{tot}$ achieves 
$\epsilon_{min}=1/{\sqrt{\left[1+4 n_{tot}\right]}}$. 
We show these results graphically in Figure (\ref{fig1}) for $N=10$.

Now how do we interpret these results. For large $n_{tot}$ (a regime we need to be 
in to achieve a good sensitivity) we find the $N$ mode entangled cat state has an 
extra critically important $\sqrt{N}$ improvement over $N$ individual copies of 
a single mode cat state (both states have the same mean photon number). 
This is due to the entangled cat state and the collective (Bell like) measurement. 
These results indicate that it would seem to be most efficient for a large fixed $n_{tot}$ to have 
as many modes in the entangled state as possible.

\section{Comparison and Discussion}
It is enlightening to compare our results to the study of Ramsey fringe interferometry
introduced by Bollinger et al. \cite{Bollinger} and discussed by Huelga et
al.\cite{Huelga}. In Ramsey fringe interferometry the objective is to
detect the relative phase difference between two superposed states,
$\{|0\rangle,|1\rangle\}$. The theory of quantum parameter 
estimation\cite{BCM} indicates in this case
that we should choose the input state as
$|\psi\rangle_i=(|0\rangle+|1\rangle)/\sqrt{2}$ and the optimal
measurement is a projective measurement in the basis
$|\pm\rangle=|0\rangle\pm|1\rangle$. The probability to obtain the result
$+$ is  $P(+|\theta)=\cos^2\theta$. In $N$ repetitions of the measurement the
uncertainty in the inferred parameter is $\delta\theta=1/{\sqrt{N}}$
which achieves the lower classical bound for quantum phase parameter estimation.
However it was first noted by Bollinger et al.\cite{Bollinger}  that
a more effective way to use the $N$ level systems is to
first prepare them in the maximally entangled state.
\begin{equation}
|\psi\rangle=\frac{1}{\sqrt{2}}(|0\rangle_1|0\rangle_2\ldots|0\rangle_N+|1\rangle
_1|1\rangle_2\ldots|1\rangle_N)
\label{entangle}
\end{equation}
In this case  parameter estimation gives the Heisenberg lower bound of $\delta\theta=1/{N}$. 
The Hilbert space of $N$ two level systems is the tensor product space of 
dimension $2^N$. The entangled state in Eq.(\ref{entangle}) however resides in a
lower dimensional subspace of permutation symmetric
states\cite{Dicke}. It can be written in the form
\begin{equation}
|\psi\rangle=\frac{1}{\sqrt{2}}(|-N/2\rangle_{N/2}+|N/2\rangle_{N/2})
\end{equation}
and so we can regard the state as an SU(2) `cat state' for $N$
two-level atoms. Hence it is straightforward to see that a single
$N$ level atom can achieve the same frequency sensitivity. 
Their equivalence can be also be understood by noting that 
the sensitivity of such frequency measurement is
proportional to the energy difference of the states involved. What
entanglement allows is for one to create an effective state 
without the need of resorting to create a superposition between 
certain ground state and a highly excited one.

{\bf To conclude}, we have in this article shown how superpositions of
coherent states can be used to achieve extremely sensitive force
detection. For a single mode state $|\alpha\rangle + |-\alpha\rangle$ we have found that the
minimum detectable displacement for weak force measurements scales as $1/\sqrt{n_{tot}}$.
More importantly by using an $N$ mode entangled cat state (and a collective measurement) a 
sensitivity scaling as $\epsilon_{min}=1/\sqrt{4 N n_{tot}}$ is achievable. This sensitivity is at the 
Heisenberg limit and indicates even with finite mean photon number $n_{tot}$ (or fixed mean energy) that 
it is possible to achieve extremely sensitive displacement detection by letting $N$ become large.

\section*{Acknowledgements}
WJM, KN and SLB acknowledges funding by the European projects
EQUIP, QUICOV and QUIPROCONE. GJM 
aknowledges support from the Australian 
Centre for Quantum Computer Technology.


\vspace{-0.2 cm}
\begin{thebibliography}{99}
\vspace{-0.2 cm}
\small


\bibitem{Huelga}  S. F. Huelga, C. Macchiavello, T. Pellizzari, A. K. Ekert,
M. B. Plenio and J. I. Cirac, Phys.Rev.Lett. {\bf 79} , 3865, (1997) .

\bibitem{Braginsky92} V. Braginsky and F. Ya. Khalili, {\em Quantum
measurement}, (Cambridge University press, Cambridge, 1992).

\bibitem{WallsMil94} D. F. Walls and G. J. Milburn, {\em Quantum Optics},
(Springer, Berlin, 1994).

\bibitem{Caves} C. M. Caves, K. S. Thorne, R. W. P. Drever, V. D. Sandberg and M.
 Zimmermann, Rev. Mod. Phys. {\bf 52}, 341, (1980).
 
\bibitem{Parkins2000} A. S. Parkins, E. Larsabal,  Phys. Rev. A 63, 012304
(2001)

 \bibitem{Bollinger} J. J . Bollinger, Wayne M. Itano, D. J. Wineland and D.
J. Heinzen, Phys. Rev. A. {\bf 54}, R4649, (1996).


\bibitem{BCM}S. L. Braunstein , C. M. Caves and G. J. Milburn, Annals of
Physics, {\bf 247}, 135, (1996).

\bibitem{Dicke} R. H. Dicke, Phys. Rev. {\bf 93}, 99 (1954).


\end{thebibliography}
\end{document}